\documentclass{ws-p8-50x6-00}

\begin{document}

\title{Physics at TESLA}

\author{G.A.Blair}

\address{Department of Physics, Royal Holloway, University of London,
Egham, Surrey. TW20 0EX, UK\\
E-mail: g.blair@rhul.ac.uk}


\maketitle

\abstracts{
The physics at a 500--800~GeV electron positron linear collider, TESLA,
is reviewed.  The machine parameters that impact directly 
on the physics are discussed and a few key performance goals
for a detector at TESLA are given. 
Emphasis is placed on precision
measurements in the Higgs and top sectors and on extrapolation
to high energy scales in the supersymmetric scenario.  
}

\section{Introduction}
A high energy linear collider (LC) operating in the energy 
regime 0.5--$\sim$ 1~TeV
will provide a programme of precision measurements and possibly new
discoveries in an energy regime matching that of the LHC.  Worldwide 
studies~\cite{WorldStudy} have been exploring the physics potential of
the LC, representing machine designs from Japan (JLC), 
America (NLC) and Europe (TESLA). A path to 
multi-TeV machines is also proposed by CERN (CLIC).
The TESLA design is well advanced and a Technical Design 
Report (TDR)~\cite{TDR}
was published in March 2001 that covers the machine~\cite{Winfred}, 
the detector and the physics.

\section{Experimental Environment}
The advantage of a linear collider (LC) over a hadronic one (LHC) 
centres on the
facts that the centre of mass (cms) energy of the colliding particles is
tunable (given by energy of the colliding beams) and that
the events themselves are much cleaner.  
The LC has the additional advantage that it can provide polarised beams, 
which can be used to explore the spin structure of fundamental
couplings, to reduce systematic errors in precision measurements,
to provide an effective luminosity gain in some channels 
such as the production of chiral supersymmetric particles 
(sparticles) and, importantly, to reduce dramatically the backgrounds to
some processes from $W$-pair production.

Both the LHC and the LC will be needed to explore fully
the TeV scale.  Some states that are invisible at the LHC, 
such as an invisible Higgs, would be
both discovered and have their properties measured at the LC.
Other states that may be discovered at the LHC will, if their mass
is not too high, be studied with high precision at the LC.  The
benefits apply in both directions; for instance if supersymmetry (SUSY)
is realised in nature then the LHC will almost certainly discover it
and will tell the LC where to concentrate its threshold scans.
However LHC analyses are limited by the overall scale, set by
the lightest SUSY particle (LSP) mass.  A precision measurement
of the LSP mass at the LC would then feed back directly to the
LHC analyses.  This type of feedback
would clearly be greatly beneficial to both programmes and is
a strong argument for concurrent running of the LHC and the LC.
\subsection{Machine Parameters}
\label{subsec:MachineParms}
TESLA will provide tunable $e^+e^-$ collisions at cms
energies ranging from the $Z^0$ pole up to about 0.8~TeV.  The machine offers
very high luminosity of 3--5$\times 10^{34}$~cm$^{-2}$s$^{-1}$.
This performance permits a general analysis to assume a total integrated
luminosity of order 500~fb$^{-1}$ and specific high-importance
analyses (such as the Higgs self coupling, discussed below) to 
assume an integrated luminosity of order 1 ab$^{-1}$.
Electron polarisation of $\sim$80$\%$ is planned and
positron polarisation of up to $\sim$60$\%$ is also envisaged
(there will be some trade-off between luminosity and positron
polarisation).  The polarisation can be measured to a 
precision of about 0.5$\%$.

Although the nominal beam energies will be known to the $10^{-4}$ level,
there are additional effects influencing the energy of individual
events.  One process, already familiar at LEP and SLC
energies, is initial state radiation.  An additional effect
at the LC is beamstrahlung, where the high fields in the 
colliding bunches cause the emission of 
large numbers of photons, some of which will in turn produce
$e^+e^-$ pairs.  Beamstrahlung results in a further widening
of the effective electron/positron energy spectrum at the level of
3.3--4.7$\%$.

The TESLA design proposes head-on collisions of electrons
with positrons in one interaction region (IR).  A second IR
is also envisaged with a finite opening angle between
the incoming beams.  This second IR could
be dedicated to $\gamma\gamma$ collisions and would need a specialised
detector and final focus layout.  In addition
to the $e^+e^-$ and $\gamma\gamma$ options, it would also
be possible to run in $e^-e^-$ or $e^-\gamma$ modes.

\subsection{A Detector for TESLA}\label{subsec:detector}
The detector at TESLA must be able to reconstruct 
high multiplicity multi-jet events as well as be able to
measure precisely high-energy single particles.  The detector
must be hermetic and the TDR design allows energy measurements down
to about 27 mrad with tagging of high energy 
electrons/positrons down to about 4~mrad.  The low angle region is
important for a range of physics processes; these include
multi-jet events where at least one jet will
be in the forward direction, SUSY processes - especially when the mass 
differences between the sparticles are small, and also for 
measuring (or vetoing) the large numbers of $\gamma\gamma$ events.  
Tracking at low angles
is also important to be able to measure the centre of mass
energy spectrum using acolinear Bhabha events.
The detector design at low angles, and in the
vertex detector region, is constrained by the 
presence of large numbers of $e^+e^-$ pairs produced by the beamstrahlung 
photons mentioned in Sec.~\ref{subsec:MachineParms}.  A higher magnetic field
is important to contain these particles at low-angle trajectories
and a 4T detector field is presently being considered.

In a typical hadronic event, the individual particle energies are typically
of order 1--2~GeV, so the quality of reconstruction hinges on the
ability to measure
very precisely the positions, directions and energies of lower-energy
particles.  The process whereby individual particles are reconstructed
by combining  information optimally from both tracking and calorimetry is
called energy flow and the goal performance is set at 
$\delta E \sim 0.3 \sqrt{E{\rm(GeV)}}$, a factor of two improvement
on that achieved at LEP/SLC.  This performance will require
very high resolution and high granularity calorimetry, coupled
with very high performance tracking.

In addition to measuring well the lower momentum tracks, it is also important
to be able to make an excellent measurement of high momentum tracks, for
instance to reconstruct the Higgs mass from recoil against the $Z^0$ boson
(see Sec.~\ref{subsec:Higgs}).  A global tracking performance of
$\delta (1/p_T)\sim 5\times 10^{-5}$~GeVc$^{-1}$ is required.
Excellent vertexing, for flavour identification, is essential to a 
wide range of physics processes, including measuring the branching ratios 
of the Higgs boson(s) and reconstructing the complicated cascade decay 
chains of SUSY.  A high performance vertex detector is also essential
to help with pattern recognition in high-density jets and to
provide additional high-precision spatial measurements, which are essential
to the global tracking performance.

\section{Physics Processes}
The physics programme at the LC is very rich; a comprehensive
overview cannot be given here and only a few highlights from the
Higgs and SUSY studies are presented in the following sections.
A whole set of interesting measurements can also be made in the top 
sector, for instance the top quark mass can be measured to a precision 
where theoretical errors dominate (at $\delta m_t \sim 100$~MeV). 
Constraining the strong coupling $\alpha_s$ to the world average would give a 
top mass precision of order 40~MeV, an order of magnitude improvement
on the LHC value~\cite{martinez}.  
High luminosity running at lower energies (the $Z^0$ pole
and the $W$-pair threshold) can lead to typically an order 
of magnitude improvement on the current measurements of electroweak 
parameters, as shown in Tab.\ref{tab:GigaZ}~\cite{TDR}.
\begin{table}[t]
\caption{Precisions on electroweak parameters from running 
TESLA at lower energies.\label{tab:GigaZ}}
\begin{center}
\begin{tabular}[c]{|c|c|c|}
\hline
 & LEP/SLC/Tevatron &  TESLA \\
\hline
$\sin^2\theta^\ell_{\rm eff}$ & $0.23146 \pm 0.00017$ & { $\pm 0.000013$}\\
\hline
\multicolumn{3}{|l|}{lineshape observables:}\\
\hline
$M_Z$ & {$ 91.1875 \pm 0.0021{~\rm GeV}$} & { {$ \pm 0.0021{~\rm GeV}$}} \\
$\alpha_s(M_{Z^2})$ & {$ 0.1183 \pm 0.0027 $} & {{$ \pm 0.0009$}} \\
$\Delta \rho_\ell$ & {$ (0.55 \pm 0.10 ) \cdot 10^{-2}$} 
& {{$ \pm 0.05\cdot 10^{-2}$}}  \\
$N_\nu$ & {$ 2.984 \pm 0.008 $} & {{$ \pm 0.004 $}} \\
\hline
\multicolumn{3}{|l|}{heavy flavours:}\\
\hline
$\cal{A}$$_b$ & $0.898 \pm 0.015$   &{$\pm 0.001$} \\
$R_b^0$ &$0.21653 \pm 0.00069$ & {$\pm 0.00014$}  \\
\hline
$M_W$ & $80.436 \pm 0.036{~\rm GeV}$ & {$\pm 0.006{~\rm GeV}$} \\
\hline
\end{tabular}
\end{center}
\end{table}
\subsection{Higgs Physics}\label{subsec:Higgs}
Understanding electroweak symmetry breaking lies at the heart 
of the LC physics
programme.  This means proving that the Higgs boson 
has spin-0 and has couplings to the standard model fermions that 
are proportional to the mass of the fermion, that
the couplings to the gauge bosons are those expected from the
standard model and that the Higgs potential has the right form, as
given by the value of its self couplings. 

The Higgs boson can be extensively studied at the LC.  
Its mass can be measured to high precision, for example
a 120~GeV standard model Higgs mass can be measured to 40~MeV
by combining all channels and to 70~MeV from using the
$Z^0$-recoil alone, a method that will work even if the Higgs 
decays invisibly.  In addition, the Higgs tri-linear self coupling 
can be measured to a statistical precision 
of about $22\%$ if its mass is 120~GeV~\cite{margarette}.

\begin{figure}[t]
\centerline{
\epsfig{
file=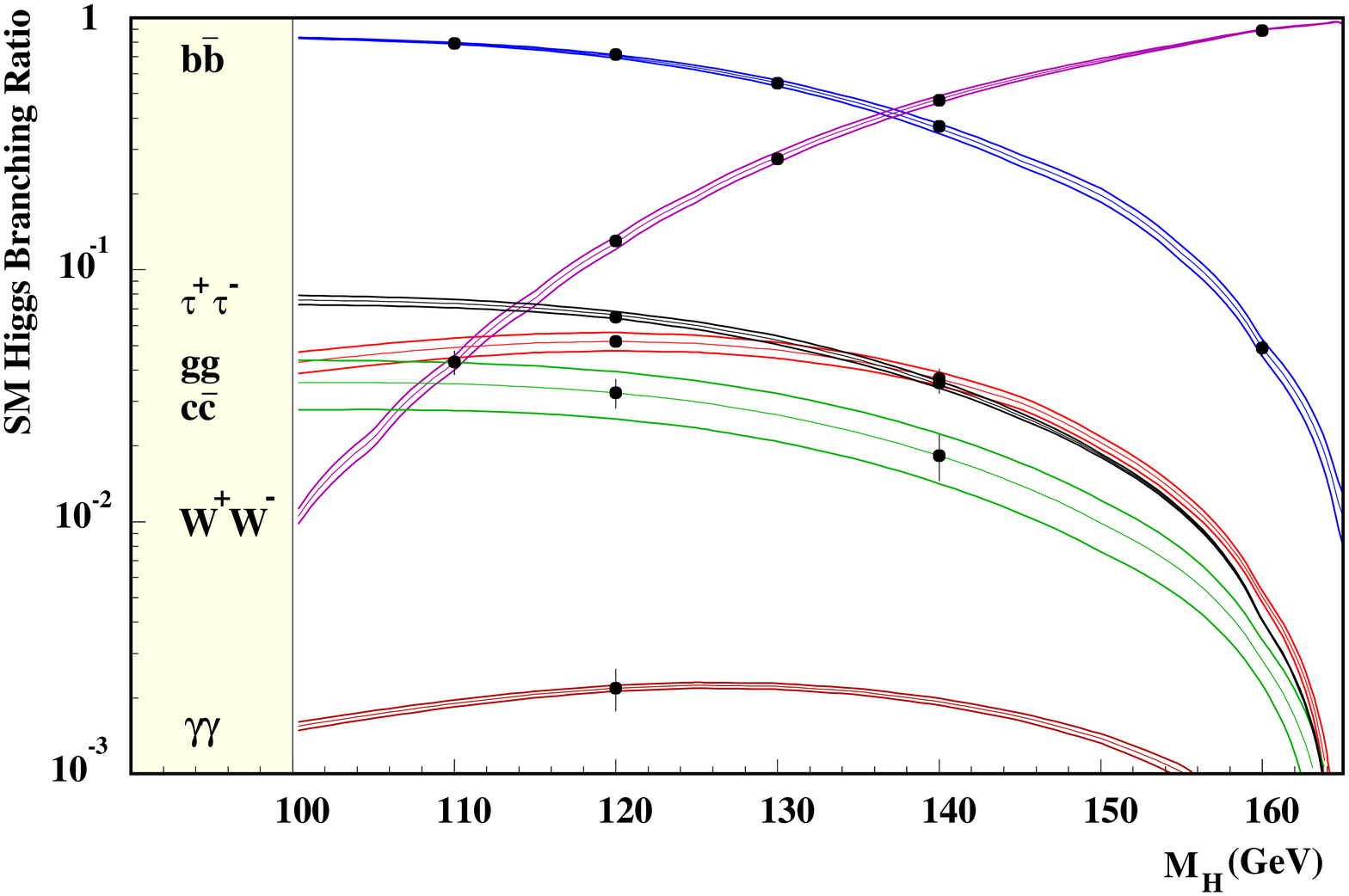, 
height=3.4in, width=3.4in}}
\vspace{10pt}
\caption{The precisions that can be obtained on the Higgs branching ratios
for a range of Higgs masses. The points are the prospective
experimental errors, the lines are the standard model 
uncertainties. 
\label{fig:higgs_br}}
\end{figure}

The Higgs spin, parity and charge conjugation quantum numbers 
can also be measured model-independently as
can its branching ratios, for which the prospects are shown in 
Fig.~\ref{fig:higgs_br}~\cite{battaglia}. 

Additional information about the Higgs can also be
obtained from increased precision measurements at the LC
of the electroweak parameters, such as those listed in Tab.~\ref{tab:GigaZ}.
For example, from such precision measurements the Higgs mass can indirectly be
constrained at the level of 5$\%$~\cite{weiglein}.  Comparing this result to
that obtained from direct measurements will provide a very
powerful test of the Higgs sector.

\subsection{Supersymmetry}\label{subsec:SUSY}
If SUSY is realised in nature, discovering it will require more than simply
observing events with missing energy/momentum that are consistent
with a given scenario.  It will require measuring the couplings
of the sparticles, testing that they are the same as those of
the particles and also that the sparticles have the correct spin.

The current data favour a light Higgs boson and SUSY provides a
mechanism that can accommodate such a light Higgs in a natural manner.
The prospects for SUSY thus look good at present and, if SUSY is indeed
realised in nature, the LC is ideal for making precision measurements of
sparticle masses and cross sections.  This is achieved by measuring the 
end-points of the energy spectra of the daughter particles 
and by performing threshold 
scans.  Using such methods and allowing 100 fb$^{-1}$ per threshold leads
typically to per-mille precision on the sparticle masses~\cite{martyn}. 

\begin{figure}[t]
\centerline{
\epsfig{
file=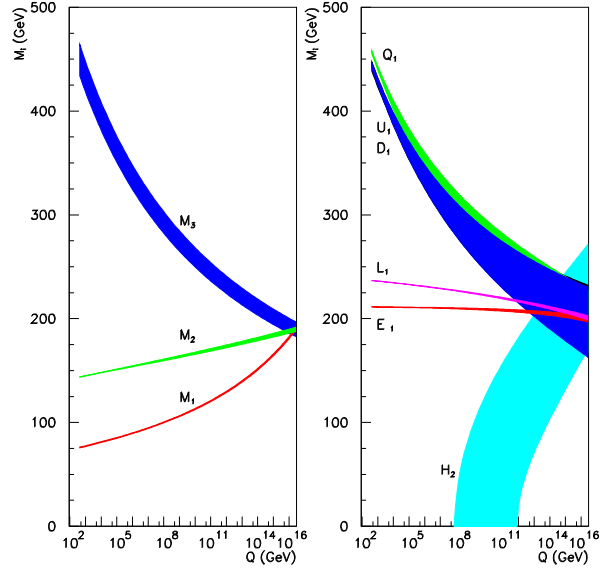, 
height=3.4in, width=3.4in}}
\vspace{10pt}
\caption{ Model-independent extrapolations
to high energies of SUSY soft breaking terms.  The left hand plot shows
the extrapolations of the gaugino mass parameters and the
right hand plot the scalar mass parameters.  The underlying
model is minimal supergravity 
with $\tan\beta=30$, $M_0=$200~GeV,$M_\frac{1}{2}=$190~GeV,
$A_0=$550~GeV, sign($\mu$)=-1.
\label{fig:susy_plots}}
\end{figure}

After measuring the masses at the TeV scale, it is then possible
to extrapolate the theory to higher energies by running the renormalisation
group equations.  In this way, the underlying structure of nature
is reconstructed in a model-independent way~\cite{BPZ}.  Applying this 
procedure to a minimal supergravity model leads to Fig.~\ref{fig:susy_plots}
where the very fine lines are obtained from the LC measurements of
the colourless sparticles and the broader ones assume LHC precisions
for the coloured states (which tend to be heavier).  In this
case, the observation that the curves meet at a common point would indicate
that gravity plays a direct role in the breaking of SUSY.

\section{Summary}
The physics and detector studies reported here are only a small
subset of a large amount of work by many people.  More
details can be found in the TDR~\cite{TDR} and in a complete
set of LC-Notes~\cite{LCNotes}, including work from all
the international regions.  It is clear that the LC presents
a very exciting physics potential and much benefit to
both the LHC and the LC programmes could be obtained 
from a significant time overlap in the running of these machines.  
The nature of electroweak symmetry breaking, the precision measurement of
(s)particle properties and detailed extrapolations to high
energy scales may yield model-independent answers to very fundamental
questions.

\section*{Acknowledgments}
Financial support from DESY and from the British Council ARC programme is 
gratefully acknowledged.

\end{document}